\documentclass[RNAAS]{aastex62}

\begin{document}

\title{The double blue straggler sequence in NGC 2173: a field contamination artefact?}

%% Note that the corresponding author command and emails has to come
%% before everything else. Also place all the emails in the \email
%% command instead of using multiple \email calls.
\correspondingauthor{Chengyuan Li}
\email{chengyuan.li@mq.edu.au}

\author[0000-0002-3084-5157]{Chengyuan Li}
\affiliation{Department of Physics and Astronomy, Macquarie University, Balaclava Road, Sydney, NSW 2109, Australia;} 
\affiliation{Department of Astronomy, China West Normal University, Nanchong 637002, People's Republic of China;}

\author[0000-0001-9073-9914]{Licai Deng}
\affiliation{School of Astronomy and Space Science, University of the Chinese Academy of Sciences, 20A Datun Road, Beijing 100012, People's Republic of China;} 
\affiliation{Department of Astronomy, China West Normal University, Nanchong 637002, People's Republic of China;}

\author[0000-0002-7203-5996]{Richard de Grijs}
\affiliation{Department of Physics and Astronomy, Macquarie University, Balaclava Road, Sydney, NSW 2109, Australia;}
\affiliation{International Space Science Institute--Beijing, 1 Nanertiao, Zhongguancun, Beijing 100190, People's Republic of China;}
\affiliation{Department of Astronomy, China West Normal University, Nanchong 637002, People's Republic of China;}

\author[0000-0003-4265-7783]{Dengkai Jiang}
\affiliation{Yunnan Observatories, Chinese Academy of Sciences, Kunming 650216, People's Republic of China;} 
\affiliation{Key Laboratory for the Structure and Evolution of Celestial Objects, Chinese Academy of Sciences, Kunming 650011, People's Republic of China;}
\affiliation{Center for Astronomical Mega-Science, Chinese Academy of Sciences, Beijing 100012, People's Republic of China;}

\author[0000-0003-4265-7783]{Yu Xin}
\affiliation{School of Astronomy and Space Science, University of the Chinese Academy of Sciences, 20A Datun Road, Beijing 100012, People's Republic of China;}

%% Note that RNAAS manuscripts DO NOT have abstracts.

%% Start the main body of the article. If no sections in the 
%% research note leave the \section call blank to make the title.

\section{Abstract} 
\cite{Li18a} (hereafter L18) detected two apparently distinct
populations of blue straggler stars (BSSs) in the young globular
cluster NGC 2173, a similar feature as observed in numerous Galactic
globular clusters (GCs). However, because of the large distance to NGC
2173, which is located in the Large Magellanic Cloud, precise proper
motion measurements for these BSSs are unattainable. In addition,
there are no observations of any nearby reference field observed with
the same instrumental setup ({\sl Hubble Space Telescope} equipped
with UVIS/WFC3 and using the F336W and F814W filters), which renders
estimating the level of field contamination for these BSSs
difficult. Recently, \cite{Dale18a} (D18) compared the observed
color--magnitude diagrams (CMDs) of both the cluster and a nearby
reference field (although observed with the ACS/WFC instrument). They
conclude that the bifurcated pattern of BSSs in NGC 2173 observed by
L18 is a field contamination artefact.

In this note, we explore the central concentration properties of the
removed `field stars' identified by D18. Our purpose is to examine if
these `field stars' are spatially homogeneously distributed. Employing
a Monte Carlo-based approach, we have carefully studied the
probability that any such central concentration may be caused by small
number statistics. We find that, in most cases of, the `field stars'
removed by D18 exhibit a clear central concentration, which cannot be
explained on the basis of small number statistics alone. Therefore, we
suggest that D18 may well have overestimated the field contamination
level, implying that the bifurcated BSS pattern in NGC 2173 cannot, in
fact, be explained by field contamination.

\section{Method}

For distant clusters without precise measurements of their stellar
proper motions, the best method to estimate field contamination by
comparing with photometric measurements of a nearby reference
field. However, L18 highlighted the unavailability of an appropriate
nearby reference field observation, which makes precisely estimating
the level of field contamination impossible. As an alternative
approach, one can examine the central concentration of the stars of
interest, combine with Mento Carlo-based simulations aimed at
quantifying whether the observed population of stars may be composed
of cluster members. This latter method has been used by many previous
studies \citep[e.g.,][]{Li18b}.

\begin{figure*}
\includegraphics[width=1\columnwidth]{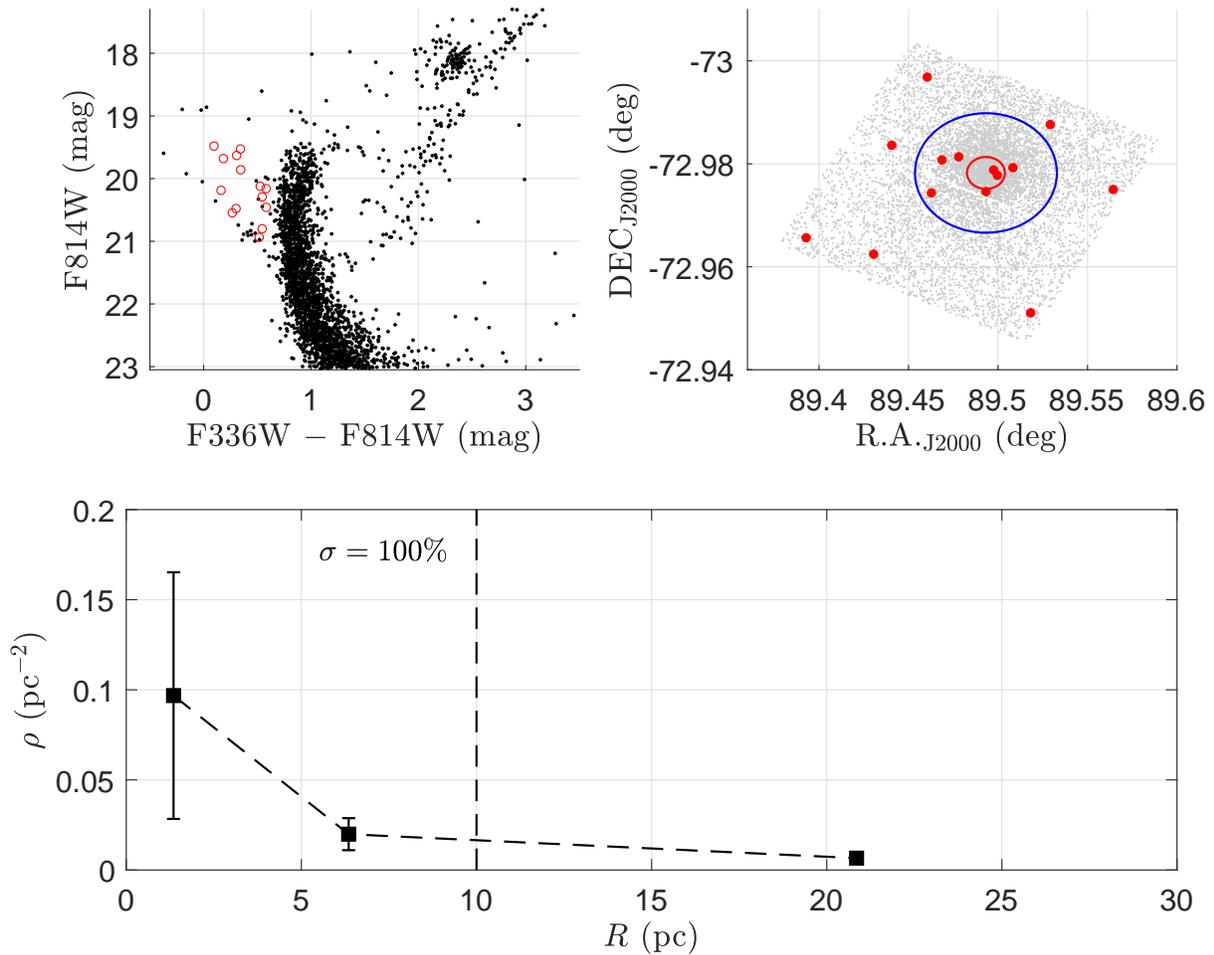}
\caption{(top left) NGC 2173 CMD. Red circles are `field stars'
  removed by D18. (top right) Spatial distribution of the D18 `field
  stars'. The blue and red circles indicate the positions of the core
  and half-mass radii, respectively (L18). (bottom) Number density
  profile of the `field stars' removed by D18. The vertical dashed
  line indicates the half-mass radius (L18). The central
  concentration's significance ($\sigma$) is almost 100\%.}
\label{F1}
\end{figure*}

By comparison of the photometric measurements of L18 and D18 for the
cluster's BSS region, we determined which `field stars' were
identified by D18 (see their Fig. 5). In the left-hand panel of
Fig. \ref{F1} we show the NGC 2173 CMD, highlighting the D18 `field
stars'. In the right-hand panel, we show their spatial distribution.

We next calculated the number density profile of these `field
stars'. If they are genuine field stars, a central peak in their
number density profile would not be expected. Because of their small
number (14), we calculated their number densities in only three radial
intervals: (1) from 0 to the cluster's core radius, $r_{\rm c} = 2.7$
pc (L18), (2) from $r_{\rm c}$ to the half-mass radius, $r_{\rm h} =
9.7$ pc (L18), and (3) beyond $r_{\rm h}$. Both $r_{\rm c}$ and
$r_{\rm h}$ are indicated in the top right-hand panel of
Fig. \ref{F1}. Fig. \ref{F1} (bottom) shows the number density profile
of the D18 `field stars'. A clear central concentration is observed.

Because the number of removed stars is small, it is possible that
their spatial concentration is caused by small number statistics. To
quantify this possibility, we employed a Monte Carlo method similar to
that used by \cite{Li18b}: we randomly generated 14 artificial stars
that were spatially homogeneously distributed. We then calculated the
ratio of their numbers inside and outside $r_{\rm h}$. We repeated
this procedure 10,000 times and recorded how often the artificial
stars achieved a higher number ratios than the observations. None of
our simulations reproduced the observed concentration, implying a
probability that the observed peak of the `field stars' could have
been caused by small number statistics of less than 0.01\%.

\begin{figure*}
\includegraphics[width=1.0\columnwidth]{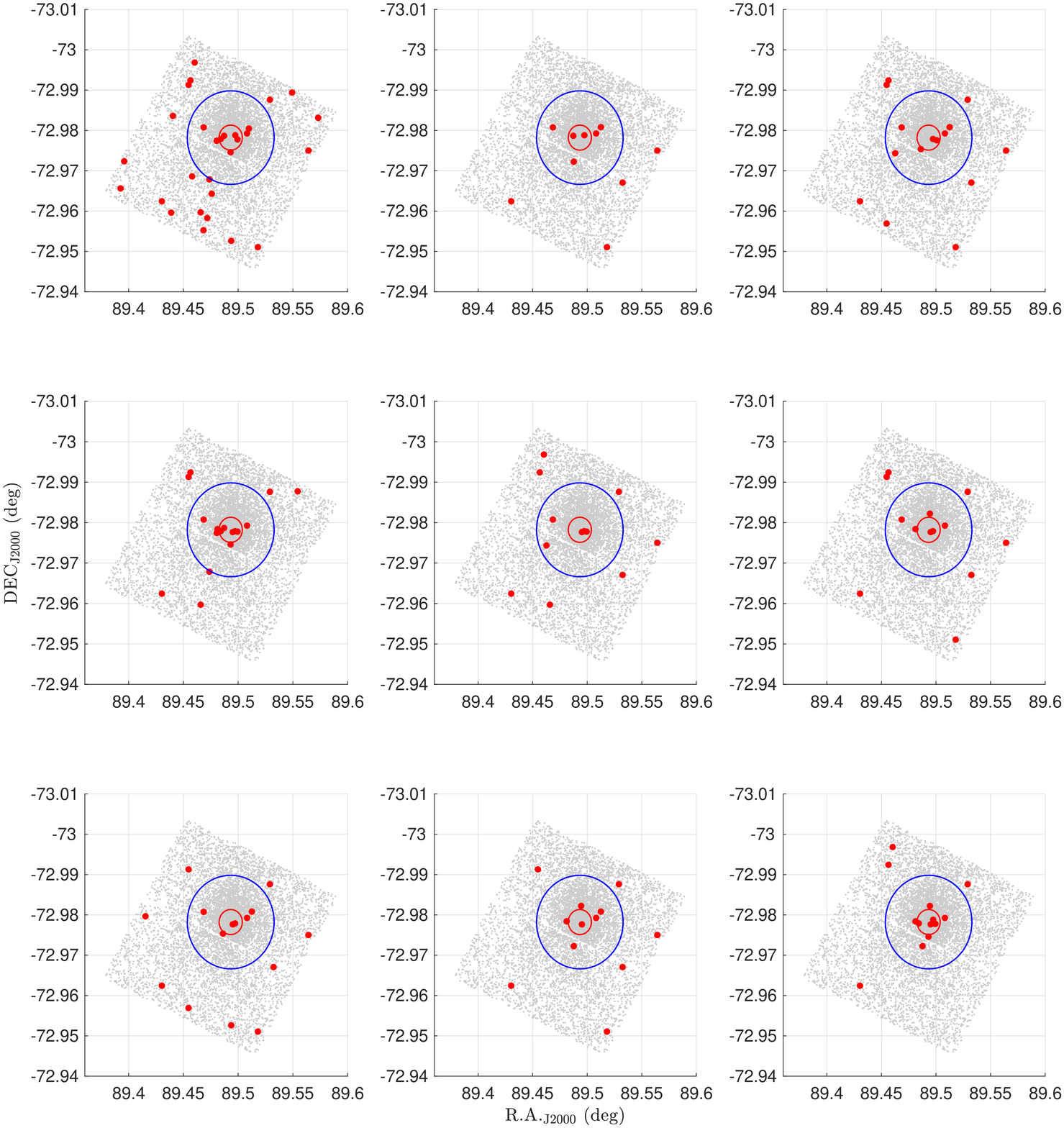}
\caption{Spatial distributions of the `field stars' removed by D18
  (their Fig. 6). The red and blue circles indicate the the core and
  half-mass radii, respectively.}
\label{F2}
\end{figure*}

In their Fig. 6, D18 show nine additional tests of `decontaminated'
CMDs, obtained by changing the color--magnitude cells' dimensions and
the grid limits, which they used for removing stars from the CMD. We
also examined the corresponding spatial distributions of the removed
`field stars' for these nine cases: see Fig. \ref{F2}. Most of these
examples again show central concentrations. In fact, the core of NGC
2173 only occupies an area of 1.6\% of the full image. If the stars
removed by D18 were field stars, in most cases we should not find any
star within the core region. However, the number ratio of stars in the
core region with respect to that outside of this region ranges from
10.7\% to 35.7\% (1 to 5 stars).

\begin{figure*}
\includegraphics[width=1.0\columnwidth]{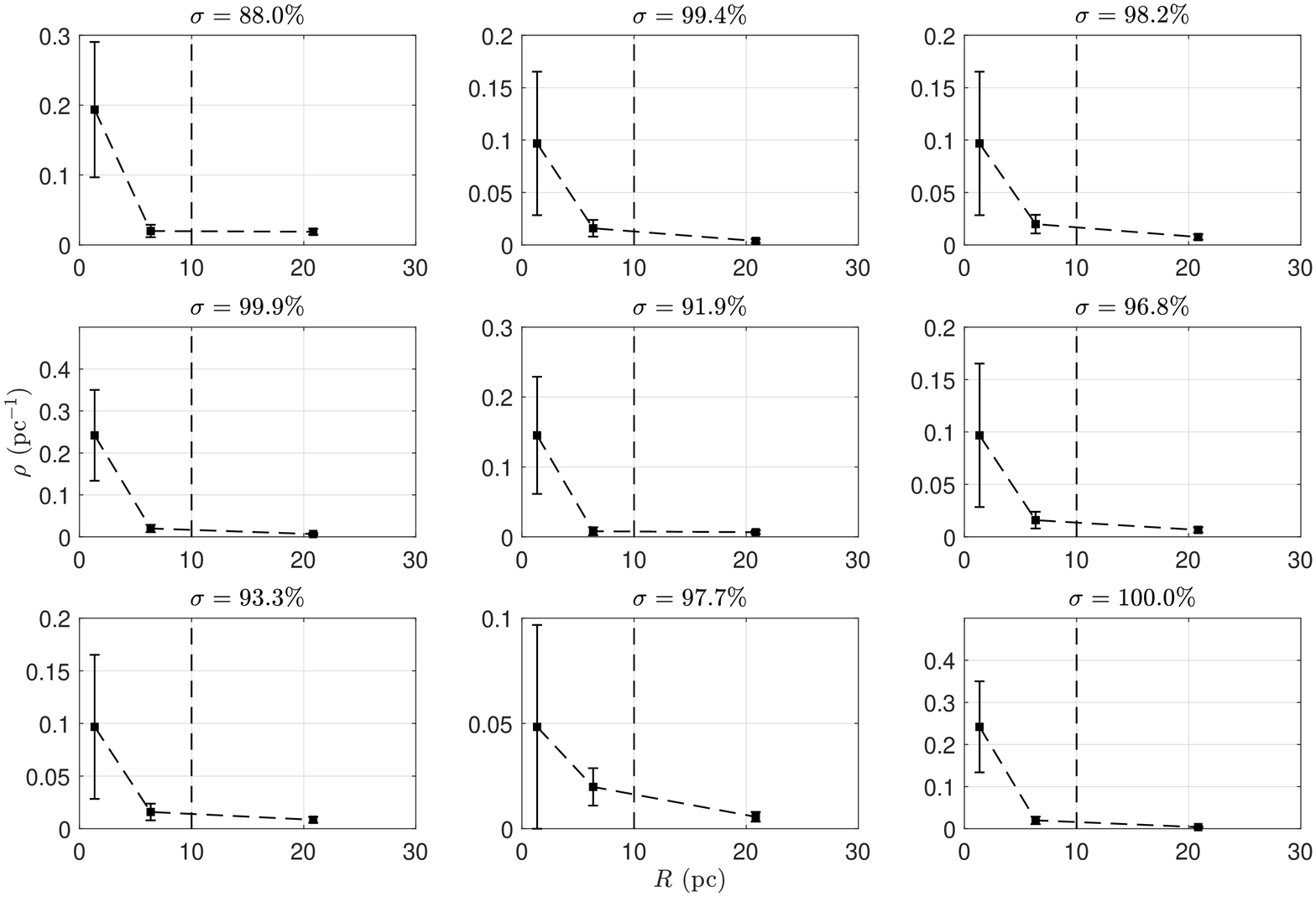}
\caption{As Fig. \ref{F2}, but the nine test cases corresponding to
  Fig. 6 of D18.}
\label{F3}
\end{figure*}

Using the same method, we plot the number density profiles and
calculate the concentrations' significance levels of the removed
`field stars' corresponding to Fig. 6 of D18: see Fig. \ref{F3}. We
find that only in one case, corresponding to the top left-hand panel
of D18's Fig. 6, the significance of the concentration is 88\%. In all
other cases, these removed `field stars' exhibit clear concentrations
with significance levels ($\sigma$) greater than 90\% (indicated at
the top of each panel in Fig. \ref{F3}). Again, these results
unequivocally suggest that D18 may have overestimated the field
contamination in all nine cases.

\section{Discussion and Conclusions}

D18 suggested that more than 40\% of the BSS candidates in the NGC
2173 CMD, may actually be field stars. As we have shown in this note,
the `field stars' removed by D18 are more likely spatially associated
with the cluster, indeed at high levels of significance. The field
contamination levels determined by D18 have most likely been
overestimated. We suggest that this disagreement with our previous
publication might have been caused by the nature of the reference
field used by D18. The observations of the cluster region and the
reference field employed by D18 are rather different: the former was
observed with the UVIS/WFC3 instrument, while the latter was observed
using the ACS/WFC. Moreover, the cluster and field regions used by D18
have been observed through different passbands and with different
exposure times. Therefore, these authors were forced to to
cross-calibrate the instrumental magnitudes difference between two
different filter systems. However, their results reveal non-zero
calibration uncertainties when one directly compares their
photometry. For instance, in Fig. 4 of D18, the slope of the
main-sequence (MS) of their reference field is clearly different from
that of the cluster, which clearly demonstrates the prevailing
calibration uncertainty. In addition, as shown in Fig. 4 of L18, the
stellar completeness for stars within $r_{\rm c}$ is lower than that
in the outer regions; the former is only $\sim$90\% of the
latter. This is not considered by D18.

In most cases, BSSs are expected to have a bimodal radial distribution
\citep[e.g.,][]{Ferr12a}, in particular in dynamically young clusters
like NGC 2173. It is possible that the average stellar number density
in the outer region would be even higher than at intermediate
radii. If so, using a BSS sample drawn from the outer region as
reference to the field, would inevitably overestimate the genuine
level of field contamination in the inner cluster
region\footnote{Although D18 state that their adopted reference field
  is located well beyond the cluster's tidal radius, both L11 and
  \cite{Mcla05a} derived a significantly larger tidal radius.}.

D18 concluded that ``NGC 2173 turns out to be a young cluster with a
single and poorly populated BSS sequence, which can likely be the
result of binary evolution'' and ``[a]s a consequence, the case of NGC
2173 is not relevant for the understanding and the discussion on the
origin of the double BSS sequences observed so far in a few old
globular clusters \citep{Ferr09a,Dale13a,Simu14a}.'' However, as we
have shown here, these conclusions are likely incorrect owing to their
overestimation of the level of field contamination. L18 estimated the
significance of the bifurcated BSS pattern in NGC 2173 at 99.95\% (see
their Fig. 5). This high significance indicates that both the blue and
red sequences are not `poorly populated'. Finally, if the bifurcated
pattern discussed by L18 is a field contamination artefact, it is
difficult to explain why field stars with different ages and
metallicities would populate two clearly distinct sequences. So far,
the most viable explanation of these two BSS sequences is that they
originated in the star cluster environment, although it is still a
mystery how one could produce such bifurcated BSS sequences in a
(dynamically) young cluster.

\end{document}